Positive selection drives faster-Z evolution in silkmoths.

Timothy B. Sackton (1*), Russell B. Corbett-Detig (1), Javaregowda Nagaraju (2), R. Lakshmi Vaishna (2), Kallare P. Arunkumar (2*), and Daniel L. Hartl (1)

1) Department of Organismic and Evolutionary Biology, Harvard University, Cambridge, MA 02144, USA

2) Centre of Excellence for Genetics and Genomics of Silkmoths, Laboratory of Molecular Genetics, Centre for DNA Fingerprinting and Diagnostics, Hyderabad 500 001, India

*to whom correspondence should be addressed: tsackton@oeb.harvard.edu or arun@cdfd.org.in

Dedication: This work is dedicated in memory of our colleague Javaregowda Nagaraju, who unexpectedly passed away during the preparation of the manuscript.

**Abstract**

Genes linked to X or Z chromosomes, which are hemizygous in the heterogametic sex, are predicted to evolve at different rates than those on autosomes. This "faster-X effect" can arise either as a consequence of hemizygosity, which leads to more efficient selection for recessive beneficial mutations in the heterogametic sex, or as a consequence of reduced effective population size of the hemizygous chromosome, which leads to increased fixation of weakly deleterious mutations due to random genetic drift. Empirical results to date have suggested that, while the overall pattern across taxa is complicated, in general systems with male-heterogamy show a faster-X effect primarily attributable to more efficient selection while the only female-heterogamy taxon studied to date (birds) shows a faster-Z effect primarily attributable to increased drift. In order to test the generality of the faster-Z pattern seen in birds, we sequenced the genome of the Lepidopteran insect *Bombyx huttoni*, a close outgroup of the domesticated silkmoth *Bombyx mori*. We show that silkmoths experience faster-Z evolution, but unlike in birds, the faster-Z effect appears to be attributable to more efficient positive selection in females. These results suggest that female-heterogamy alone is unlikely to be sufficient to explain the reduced efficacy of selection on the bird Z chromosome. Instead, it is likely that a combination of patterns of dosage compensation and overall effective population size, among other factors, influence patterns of faster-Z evolution.

**Introduction**

Sex chromosomes share several properties that lead to unique evolutionary consequences. Most notably, the hemizygosity of sex chromosomes in the heterogametic sex significantly affects rates and patterns of evolution in ways that can shed light on the relative importance of drift and selection (Bachtrog et al., 2011; Ellegren, 2011; Vicoso and Charlesworth, 2006). To the extent that beneficial mutations are on average partially recessive, the hemizygosity of the X chromosome in males will increase the efficacy of selection and lead to a faster rate of fixation of beneficial mutations relative to autosomes, as recessive mutations on the X will be immediately exposed to selection in males (Charlesworth et al., 1987; Vicoso and Charlesworth, 2006, 2009). Similarly, recessive or partially recessive deleterious mutations on the X will be more efficiently purged from the population (Charlesworth et al., 1987; Vicoso and Charlesworth, 2009). Together, these results suggest that hemizygosity should increase the efficacy of natural selection on the X for mutations that are at least partially recessive, with the effect on rates of X:A evolution determined by the relative contribution of adaptive and deleterious mutations to divergence. In species (such as many insects) where recombination is absent from the hemizygous sex, genes on the X or Z chromosome will also experience a higher effective recombination rate, and will therefore be less subject to Hill-Robertson interference effects, further increasing the efficacy of selection (Campos et al., 2013; Charlesworth, 2012).

Hemizygosity of the X chromosome also reduces its effective population size ($N_e$) relative to autosomes, because on average there are only 3 copies of the X for every 4 copies of the autosomes in a diploid population with equal numbers of breeding males and breeding females. The reduced $N_e$ of X chromosomes reduces the efficacy of natural selection, and thus a higher fraction of weakly deleterious alleles can drift to fixation on the hemizygous chromosome than on the autosomes (Vicoso and Charlesworth, 2009). However, sexual selection and differential variance in reproductive success between males and females can cause departures from equal effective numbers of breeding males and breeding females (Evans and Charlesworth, 2013). Thus, in natural populations the ratio of effective population size on the X ($N_eX$) to the autosomes ($N_eA$) is often not equal to the expected value of 0.75 (Mank et al., 2010b; Singh et al., 2007), with significant consequences for the predicted effects of hemizygosity on rates of X:A evolution (Mank et al., 2010b; Vicoso and Charlesworth, 2009).

Although these two effects – increased efficacy of selection due to partial recessivity (and in some cases higher effective recombination rates) and increased fixations by drift due to reduced $N_e$ – are opposite in cause, the empirical pattern they produce is in many respects identical: a faster-X effect, in which genes on the X chromosome have a higher rate of molecular evolution than genes on the autosomes, at least under certain conditions regarding recessivity and the amount and architecture of adaptive evolution (Connallon et al., 2012; Vicoso and Charlesworth, 2009). However, these two effects make different predictions on how faster-X (and faster-Z) effects should interact with sex-specific patterns of expression. While reduced $N_eX:N_eA$ (or $N_eZ:N_eA$) is predicted to increase fixation of deleterious alleles due to drift for all expression classes, the effects of hemizygosity in increasing the efficacy of selection for beneficial alleles should only apply when the gene in question is expressed in the heterogametic sex, and should be especially pronounced when the gene is uniquely expressed in the heterogametic sex (Baines et al., 2008; Grath and Parsch, 2012; Mank et al., 2010a).

Empirical results to date present a complicated picture, but a few broad trends emerge. In *Drosophila* and mammals, both male-heterogametic taxa with, in general, $N_eX:N_eA$ ratios equal to or greater than 0.75 (Mank et al., 2010b), male-biased genes show a strong pattern of faster-X evolution (Baines et al., 2008; Grath and Parsch, 2012; Khaitovich et al., 2005; Torgerson and Singh, 2006, 2003; Xu et al., 2012) suggesting that more efficient fixation of beneficial alleles plays a role in driving faster-X evolution for at least this subset of genes. Additionally, there is good evidence for increased efficacy of purifying selection on the X chromosome of *Drosophila* (Mank et al., 2010b; Singh et al., 2008) and inferred lower rates of fixation of weakly deleterious mutations in proteins (Mank et al., 2010b). However, overall patterns of faster-X evolution are often complex and lineage-specific (Baines and Harr, 2007; Begun et al., 2007; Connallon, 2007; Hu et al., 2013; Hvilsom et al., 2012; Langley et al., 2012; Mackay et al., 2012; Singh et al., 2008; Thornton et al., 2006; Xu et al., 2012) and depend on lineage-specific details regarding the relative proportions of fixations due to beneficial and weakly deleterious mutations, as well as differences in $N_e$ (Mank et al., 2010b) and lineage-specific variation in male-mutation bias (Xu et al., 2012).

Birds, where female heterogamy (females are ZW and males are ZZ) is predicted to lead to a faster-Z effect, present a very different picture: faster-Z evolution in this clade appears to be largely a function of increased fixation of weakly deleterious alleles, driven by very low $N_eZ:N_eA$ ratios that are significantly below 0.75 (Mank et al., 2010a). Under these conditions, which may be common to many female-heterogametic (ZW) taxa, the consequences of low $N_eZ$ appear to outweigh the consequences of hemizygosity, leading to less efficient selection. Crucially, this distinction between XY and ZW taxa – more efficient selection on the X, less efficient selection on the Z – relies on a single observation of rates of evolution on the Z chromosome in birds (Mank et al., 2010a, 2007). In order to better understand general patterns of sex chromosome evolution, data from additional female-heterogametic taxa are critical.

Here, we present the genome sequence of *Bombyx huttoni*, a close relative of the domesticated silkmoth *Bombyx mori*, and use this genome sequence to analyze faster-Z evolution in silkmoths (Lepidoptera). This is to our knowledge the first analysis of faster-Z evolution in any taxa besides birds. We first show that our *B. huttoni* assembly provides more than adequate coverage for molecular evolutionary studies.

Comparing both dN/dS ratios and estimates of selection derived from published polymorphism data across expression classes (male-biased, female-biased, and unbiased) indicates a strong faster-Z effect for female-biased genes, an intermediate faster-Z effect for unbiased genes, and no faster-Z effect for male-biased genes. This contrasts with the pattern observed in birds (equal faster-Z effect across all expression classes) and suggests that more efficient selection may be driving the faster-Z effect in silkmoths despite an estimate of $N_eZ:N_eA$ significantly below 0.75. We propose that conditions under which drift can predominate in sex chromosome evolution may be relatively uncommon, even in female-heterogametic taxa.

**Results**

**Assembly of the *B. huttoni* genome.**

In order to compare rates of evolution on the silkmoth Z chromosome and autosomes, we sequenced the genome of *B. huttoni*, a close outgroup to the domesticated silkmoth *B. mori (Arunkumar et al., 2006)*, by Illumina sequencing. We generated 52 million 100bp paired end reads from *B. huttoni* samples, which we assembled using a *de novo* assembly pipeline (see methods for details). The final assembly consists of 1,134,960 scaffolds and unscaffolded contigs (for linguistic simplicity, we refer to both of these as contigs even though some of them are scaffolded based on the paired-end sequencing) of at least 100 bp, containing 507.9 Mb of sequence, and with an N50 = 731 bp (Figure 1A). This represents a rather fragmentary genome, likely due to the short insert size of the sequencing library and polymorphism among the individuals used to prepare the DNA for sequencing.

Despite the fragmented nature of *B. huttoni* genome, we are able to recover orthologous sequence to a large fraction of genes in the *B. mori* reference (International Silkworm Genome Consortium, 2008). We initially used promer (Kurtz et al., 2004) to map all 1.1 million contigs to the *B. mori* reference set of protein-coding genes, and then filtered the output to retain only 1-to-1 mappings (see methods). Based on these initial 1-to-1 mappings, we have at least some *B. huttoni* sequence for 12,842 of the 13,789 genes in *B. mori* with a chromosomal location (93.1%), with a median coverage of 81.6% (including those with no coverage; Figure 1B). Just over 1/3$^{rd}$ of all *B. mori* genes (36.7%) are fully or almost fully covered (>90%) by *B. huttoni* sequence, and only 7.4% have below 25% coverage. Overall, then, despite the fragmentary nature of our draft genome, we have easily sufficient coverage of genes to estimate genome-wide evolutionary parameters.

After our initial promer mapping, we refined alignments using the alignment program FSA (Bradley et al., 2009) and custom perl scripts (see methods for details). After refinement, these 12,842 alignments contain 13.2 Mb of coding sequence, which represents 79.2% of all protein coding bases localized to chromosomes in *B. mori*. There is no difference between Z-linked and autosomal genes in either the proportion of covered bases (Z = 0.795, A = 0.792) or the fraction of genes with alignments (Z = 0.940, A = 0.934). We also compared the read coverage on the Z and the autosomes using unique mappings between contigs and the RepeatMasked *B. mori* reference sequence generated using nucmer (Kurtz et al., 2004; see methods for details). Based on these unique mappings, we computed weighted mean coverage (using contig length as the weight) for both the Z and all autosomes, and tested for a difference in coverage using a weighted T test. The Z chromosome has slightly higher weighted coverage (23.3x) compared to autosomes (21.3x), a difference that is highly significant ($P < 2.2 \times 10^{-16}$, weighted T test). We then filtered this alignment set to remove alignments with either low coverage, short length, or premature stop codons (see methods for details), leaving us with a final total of 10,639 gene alignments containing 9.93 Mb of aligned sequence to analyze.

**Faster Z evolution in silkmoths.**

Based on the 10,639 *B. huttoni*/*B. mori* alignments described above, we calculated pairwise ω, dN and dS using maximum likelihood methods in PAML version 4.4d. Overall rates of divergence are moderate, with median dS = 0.271 and median dN = 0.0219. The genes on the Z chromosome evolve more rapidly than autosomes (median ω for autosomes = 0.0785, for Z = 0.0977, P = 3 x $10^{-5}$, Wilcox-Mann-Whitney test). This pattern holds for dN as well (Z = 0.0257, A = 0.0218, P = 7.3 x $10^{-4}$), but not for dS (Z = 0.2676, A = 0.2708, P = 0.1545) (Figure 2). If the faster-Z effect is primarily driven by the increased efficacy of positive selection on recessive mutations in females, we expect that it will be absent in genes that are predominantly expressed in males. Conversely, we expect the faster-Z effect to be particularly strong for genes that are primarily expressed in females, as these will mostly be expressed in a hemizygous state. To test whether patterns of molecular evolution depend on patterns of sex-bias in expression, we used published microarray data from *B. mori* (Walters and Hardcastle, 2011; Zha et al., 2009) to define male-biased, female-biased, and unbiased genes (see methods for details). Sex-biased genes represent 42.7% of genes for which we have reliable expression data, including 1,402 female-biased genes and 2,067 male-biased genes, of which 59 and 135 (respectively) are Z-linked.

Consistent with the hypothesis that faster-Z evolution in silkmoths is driven by more efficient positive selection in the hemizygous sex, we find that the faster-Z effect (defined as the ratio of median ω on the Z to median ω on the autosomes) is completely absent in male-biased genes, intermediate in unbiased genes, and strongest in female-biased genes (Figure 3). For male-biased genes we find no evidence for a significant faster-Z effect (median ω for autosomes = 0.0818, median ω for Z chromosome = 0.0912, Wilcox-Mann-Whitney test P-value = 0.3295). In contrast, both female-biased (A = 0.0718, Z = 0.1107, Wilcox-Mann-Whitney test P-value = 0.00224) and unbiased (A = 0.0731, Z = 0.1052, Wilcox-Mann-Whitney test P-value = 0.00224) genes have strong statistical support for a faster-Z effect. The pattern is identical for dN (male-biased P = 0.74, female-biased P = 0.0033, unbiased P = 0.00435), but no class of genes shows a faster-Z effect for dS (all P > 0.05).

Focusing on the contrast between the faster-Z effect for male-biased genes and female-biased genes, we directly tested the prediction that this effect should be significantly larger in female-biased genes than in male-biased genes using a permutation test (see methods for details). The observed difference in faster-Z effect between female-biased and male-biased genes (0.475) is significantly larger than expected under the null hypothesis (one-tailed permutation P = 0.0124). Combined with the lack of statistical support for a faster-Z effect among male-biased genes, this strongly suggests that female-biased genes are qualitatively different that male-biased in the evolutionary regime they experience on the Z chromosome and provides support for the more efficient positive selection model of faster-Z evolution.

**Faster-Z evolution is due to increased rates of adaptive evolution**

An alternate approach to distinguishing more efficient positive selection from less efficient purifying selection as a cause of faster-Z evolution is to use polymorphism data to estimate the direction of selection on each gene (McDonald and Kreitman, 1991). To do this, we aligned publicly available sequencing reads from 11 strains of *B. mandarina* (Xia et al., 2009) to the *B. mori* reference, and calculated synonymous and nonsynonymous polymorphisms within *B. mandarina* and fixed differences to *B. huttoni* (using the *B. mori* sequence only as an alignment reference). Based on these polymorphism and divergence tables, we can calculate the DoS statistic, which is related to the Neutrality Index (Rand and Kann, 1996) but less sensitive to small sample sizes (Stoletzki and Eyre-Walker, 2010). DoS measures the difference in the proportion of fixed differences that are nonsynonymous and the proportion of polymorphisms that are nonsynonymous. If this statistic is positive (a higher fraction of fixed differences are nonsynonymous than polymorphisms), it implies fixation of beneficial alleles, whereas a negative statistic implies accumulation of mildly deleterious alleles.

As in the divergence data, we find that there is an overall faster-Z effect: median DoS is significantly higher for Z-linked genes than autosomal genes (Z = 0.106, A = 0.0556, Wilcox-Mann-Whitney P-value < $2.2 \times 10^{-16}$), suggesting more fixation of beneficial alleles on the Z chromosome. Consistent with the hypothesis that this overall effect is primarily driven by more efficient positive selection in females, median Z DoS is significantly greater than median A DoS for female-biased genes (Z = 0.1316, A = 0.0441, P = 0.0098) and unbiased genes (Z = 0.1138, A = 0.0585, P < $2.2 \times 10^{-16}$), but not male-biased genes (Z = 0.0909, A = 0.0646, P = 0.1012) (Figure 3). As for $\omega$, the observed difference in faster-Z effect in female-biased is significantly greater than the (non-significant) faster-Z effect for male-biased genes, based on a permutation test (one-tailed permutation P-value = 0.005).

**Variation in gene content between the Z and the autosomes**

A complicating factor in patterns of faster-Z (or faster-X) evolution is that gene content is often different between sex chromosomes and autosomes, and in particular male-biased genes are often distributed differentially between autosomes and sex chromosomes (Arunkumar et al., 2009; Ellegren, 2011; Parisi et al., 2003; Walters and Hardcastle, 2011). In at least some cases, differential gene content on the sex chromosomes can account for genome-wide faster-X effects (e.g., Hu et al., 2013); this may especially be the case to the extent that male-biased genes experience more adaptive evolution than other genes (Baines et al., 2008; Haerty et al., 2007; Meisel, 2011; Parsch and Ellegren, 2013; Pröschel et al., 2006; Zhang et al., 2004).

We find, as has been previously reported (Arunkumar et al., 2009; Suetsugu et al., 2013), that male-biased genes are overrepresented on the Z, and female-biased are depleted on the Z, relative to autosomes ($\chi^2$ P-value = $4.45 \times 10^{-5}$). As male-biased genes also evolve more rapidly overall than female-biased or unbiased genes (male $\omega$ = 0.082, other genes $\omega$ = 0.0739, Wilcox-Mann-Whitney P-value = $2 \times 10^{-5}$), in principle the overrepresentation of male-biased genes on the Z could drive a faster-Z effect. Notably, however, and consistent with the predictions of a faster-Z effect driven by more efficient selection, we do not find a faster-Z effect for male-biased genes either in $\omega$ or in DoS. This suggests that the excess of male-biased genes on the Z chromosome is not driving the faster-Z effect we observe.

**Codon bias in silkmoths**

In *Drosophila*, genes on the X chromosome exhibit significantly more codon bias than genes on the autosomes (Singh et al., 2008), which has been taken as an indication of more efficient purifying on the X chromosome due to the combination of a high $N_eX:N_eA$ ratio and more efficient selection in males (Singh et al., 2008, but see Campos et al., 2013 for an alternative explanation). In silkmoths, however, we see no difference in codon usage bias between the Z chromosome and autosomes (Z = 52.42, A =

52.93, Wilcox-Mann-Whitney P = 0.5154), after accounting for background non-coding genome composition, as measured by the corrected effective number of codons (Ncp) (Novembre, 2002; Wright, 1990), although previous reports have suggested reduced codon usage bias on the Z based on the uncorrected effective number of codons (Pease and Hahn, 2012).

**Discussion**

The unique properties of sex chromosomes are predicted to have significant effects on the evolution of sex-linked genes, which has led to numerous studies of patterns of evolution on X chromosomes relative to autosomes in several taxa, as well as limited studies of the Z chromosome of birds (Mank et al., 2010b; Vicoso and Charlesworth, 2006). Overall, a complicated picture has emerged from these results, but some general patterns are discernible. In XY taxa, the evidence for faster-X evolution of male-biased genes appears to be quite robust (Baines et al., 2008; Grath and Parsch, 2012; Khaitovich et al., 2005; Torgerson and Singh, 2006, 2003), suggesting that at least for this class of genes adaptive mutations are sufficiently common and sufficiently recessive for the predicted more efficacious positive selection on the X chromosome to lead to a faster-X effect. Beyond male-biased genes, faster-X effects are less consistent and lineage-dependent to a great degree (Baines and Harr, 2007; Begun et al., 2007; Connallon, 2007; Hu et al., 2013; Hvilsom et al., 2012; Langley et al., 2012; Mackay et al., 2012; Singh et al., 2008; Thornton et al., 2006; Xu et al., 2012). This pattern might be expected in cases where relatively high $N_eX:N_eA$ ratios due to greater variance in male reproductive success reduce or eliminate the drift-promoting effects of hemizygosity and lead to a situation where the balance of rates of positive and negative selection determine whether faster-X effects are observed (since more efficient selection on the X will increase rates of positive selection but reduce fixations of weakly deleterious mutations).

In birds – the only female-heterogametic taxon studied to date – a strikingly different pattern emerges. While there is clear evidence for a faster-Z effect in this taxon, it appears to be the result of reduced efficacy of selection on the Z, as a consequence of severely reduced $N_eZ:N_eA$ ratios, attributable to the effects on sexual selection on males in female-heterogametic taxa (Mank et al., 2010a, 2007). This observation raises the obvious question: is this a general pattern of female-heterogametic taxa, or is this result restricted to birds?

To begin to address this question of generality, we sequenced the genome of *B. huttoni*, a close outgroup of *B. mori*, the domesticated silk moth, and examined patterns of Z chromosome evolution in a lepidopteran insect for the first time. We find that Z-linked genes evolve faster than autosomal genes, but unlike results in birds this higher rate of evolution is primarily driven by a strong faster-Z effect in female-biased genes. Thus, in silkmoths the pattern of faster-Z evolution appears to be more similar to XY taxa; specifically, in silkmoths it appears that faster-Z evolution is substantially driven by more efficient positive selection on the hemizygous chromosome. This is in contrast to birds, where drift appears to predominate.

A key parameter is the relative $N_e$ of the sex chromosome to the autosomes. We estimated this parameter for silkmoths from the ratio of $\pi_A$ to $\pi_Z$ at fourfold degenerate sites in *B. mandarina* (see methods), under the assumption that this ratio has not changed dramatically between *B. mandarina* and *B. huttoni*, as 0.66 (95% bootstrap confidence interval: 0.64 – 0.68). This is somewhat higher than in the bird species that have been studied, where estimates range from 0.30 to 0.51 (Mank et al., 2010b), but it is still significantly below the expected value of 0.75. Given the absence of recombination in female Lepidoptera, which implies a higher effective recombination rate for the Z than the autosomes and thus smaller reductions in $N_e$ for neutral sites on the Z than the autosomes due to background selection (Charlesworth, 2012a, 2012b), observing such a low value of $N_eZ:N_eA$ is somewhat unexpected. However, the large number of chromosomes and relatively small sizes of each render this effect unimportant, leading to a prediction of equal effects of background selection on the Z and the autosomes (see Appendix for details), and suggesting that sexual selection likely plays a role in reducing $N_eZ:N_eA$.

Based on the numerical integrations of Vicoso and Charlesworth (2009), this value of $N_eZ:N_eA$ puts silkmoths in the region of parameter space where fixation of deleterious mutations should be elevated on the Z chromosome for all ranges of dominance, but fixation of advantageous mutations will only be elevated for relatively restrictive ranges of dominance. Qualitatively, the patterns expected for a clade with $N_eZ:N_eA$ at 0.66 and at 0.45 are not very different, and so the somewhat higher $N_eZ:N_eA$ ratio does not seem sufficient on its own to explain the difference between patterns of faster-Z evolution in birds and silkmoths. However, we cannot rule out the possibility that there is a discontinuous effect not captured in the numerical model, which produces a qualitatively different pattern of Z chromosome evolution once $N_eZ:N_eA$ falls below some threshold value.

One potential difference between birds and silkmoths is the degree to which the Z chromosome is dosage compensated. While it appears that the Z chromosome is not globally dosage-compensated in birds (Mank, 2009), the degree to which silkmoths lack of dosage compensation is unclear (Walters and Hardcastle, 2011). Lack of dosage compensation is predicted to enhance the faster-Z effect for deleterious mutations but reduce it for advantageous mutations (Charlesworth et al., 1987; Mank et al., 2010b). The reduced degree of dosage compensation in birds compared to silkmoths could thus bias the observed faster-Z effect in birds towards increased fixation of mildly deleterious alleles, as we observe.

A second likely difference between the population genetic environments of birds and silkmoths is overall $N_e$, which can have substantial consequences for patterns of sex chromosome evolution (Mank et al., 2010b; Vicoso and Charlesworth, 2009). Overall $N_e$ in the species of birds studied for faster-Z evolution is probably in the range of 200,000 – 600,000 (Mank et al., 2010b), although these estimates have a large error. In silkmoths, diversity data on the autosomes is roughly consistent with that observed in cosmopolitan *Drosophila* species (Langley et al., 2012; Xia et al., 2009), which implies an effective population size on the order of millions (assuming similar mutation rates). Thus, it is reasonable to assume that silkmoths have a higher $N_e$ in general than birds. Populations with larger $N_e$ will experience a higher rate of input of new mutations, and fewer of those new mutations will have fitness effects in the nearly neutral range. Low $N_eZ:N_eA$ ratios, which increase the fixation of mildly deleterious alleles due to drift, may thus have smaller consequences for deleterious mutations in large populations. Conversely, increased rates of adaptive evolution in large populations will disproportionally affect rates of fixation on the Z, assuming most new mutations are at least partially recessive and new mutations (as opposed to standing variation) are the source of a significant fraction of adaptive fixations. These results are consistent with the pattern we observe, in which the drift effects of hemizygosity are stronger than the selective effects in birds but the converse is true in silkmoths.

Taken together, our results suggest that female heterogamy alone may not be sufficient to explain the discrepancy observed between faster-Z evolution in birds and faster-X evolution in mammals and *Drosophila*. Instead, a combination of several factors, including the ratio of effective population size of the hemizygous chromosome to autosomes, overall effective population size, and dosage compensation likely interact to produce the patterns of sex chromosome evolution we observe across taxa. Additional studies of a more diverse array of species will help clarify the role of these forces in faster-Z and faster-X evolution.

## Methods

### Sequencing of *B. huttoni*

*B. huttoni* (also cited by the junior synonym *Theophila religiosa* in the literature) is the closest outgroup to the clade containing the domesticated silkmoth *B. mori* and its wild progenitor, *B. mandarina* (Arunkumar et al., 2006). Live pupae of *B. huttoni* were collected from their natural habitat in Northeastern India (Kalimpong, West Bengal). Genomic DNA extracted from pooled males was used for sequencing. We performed 2X100bp paired end sequencing of a genomic library of insert size 300-400bp, on an Illumina HiSeq2000 machine, using standard protocols.

### Initial *de novo* assembly of the *B. huttoni* genome

To generate the initial *de novo* assembly of the *B. huttoni* genome, we first assembled all reads using SOAPdenovo 2.04 (Luo et al., 2012), with the following options: pregraph -R -K 23 -p 48 -d 2; contig -R -M 2 -m 55 -E -p 48; map -f -k 25 -p 48; scaff -F -w -G 100 -N 500 -p 48; GapCloser -t 48 -p 25. This set of command line options implements the multi-k version of SOAPdenovo2, which uses an iterative approach to build a *de novo* assembly using k-mers of many sizes (Peng et al., 2012). After closing gaps, our initial assembly consisted of 288,089 scaffolds and 1,079,294 unscaffolded contigs, with a minimum length of 100 bp and an N50 of 680 bp.

To improve our assembly prior to analysis, we first computed average coverage for each sequence in the initial assembly (based on mapping all reads back to our assembly as described below, and then using bedtools genomecov to compute coverage) and filtered sequences with average read coverage below 5x. This eliminated 232,102 unscaffolded contigs and 858 scaffolds; we then further filtered our assembly using the REAPR pipeline (Hunt et al., 2013), which uses discrepancies in the fragment coverage distribution to detect and break misjoined scaffolds and fix related assembly problems. We implemented REAPR with default settings, including using SMALT to map reads to our assembly (we use the same mapping to compute coverage). The final assembly includes a total of 287,768 scaffolds and 847,192 unscaffolded contigs, containing 507.9 MB of assembled sequence with an overall N50 of 731 bp.

As a second quality control check on our assembly, we used nucmer (with options -maxmatch -g 1000; Kurtz et al., 2004) to map our *B. huttoni* assembly to a repeat-masked version of the *B. mori* genome version 2.3 (International Silkworm Genome Consortium, 2008), created using RepeatMasker (http://www.repeatmasker.org/) and the B. mori specific TE library available from KAIKOBase (http://sgp.dna.affrc.go.jp/data/BmTELib-080930.txt.gz). We filtered the nucmer output to identify the single best location where each query hit (contig or scaffold) maps in the reference genome, and then computed the fraction of scaffolds and contigs with hits to more than one genomic region (suggesting either false joins or genome rearrangements). Only 0.36% of contigs that align to *B. mori* map to more than genomic location, and only 10.48% of scaffolds that align to *B. mori* map to more than one genomic location.

### Mapping to *B. mori*.

In order to estimate patterns of gene evolution, we focused on generating a high-quality alignment of our *B. huttoni* assembly to *B. mori* protein-coding genes. Because both the *B. mori* genome and our fragmentary draft *B. huttoni* genome are highly repetitive in non-coding regions, the most straightforward approach is to align our *B. huttoni* assembly to *B. mori* protein-coding sequence only. To do this, we used promer (with options --maxmatch -b 150 -c 15 -g 25) to map our final assembly to the consensus gene set for *B. mori*, dated Apr-2008 and available at KAIKObase (http://sgp.dna.affrc.go.jp/pubdata/genomicsequences.html). We then filtered the resulting delta file output to retain only 1-to-1 mappings (option -1).

**Realigning *B. mori* and *B. huttoni* sequences and estimate molecular evolutionary parameters.**

In order to improve the quality of the initial promer alignments, above, we first trimmed or extended each hit between *B. mori* and *B. huttoni* to extract a single homologous exon for each promer match. We then realigned the extracted *B. huttoni* sequence to *B. mori* using FSA (Bradley et al., 2009), which is a protein-aware statistical aligner that imposes penalties for introduced frameshifts and stop codons in coding sequence. Finally, we refined the FSA alignments to fix three common errors: first, we optimized gaps to prefer terminal gaps to internal gaps; second, we trimmed *B. huttoni* sequence at alignment ends to remove low-scored regions; and third, we removed putative intronic sequence in *B. huttoni* by removing long stretches of sequence in *B. huttoni* that are aligned to gaps in *B. mori*.

After these refinement steps, we screened the remaining alignments to remove alignments with either too low coverage (defined as either fewer than 60 aligned bases or less than 10% coverage) or with premature stop codons. Of the 12,842 genes with at least some *B. huttoni* coverage, we filter 83 for coverage reasons and 2,120 due to presence of non-terminal stop codons, leaving 10,639 alignments for analysis. We then used the filtered set of FSA alignments as input to PAML 4.4d (Yang, 2007) for analysis of patterns of molecular evolution on a per-gene basis, fitting a model with one ω ratio per gene in PAML, and retained for analysis maximum likelihood estimates of dN, dS, ω, and total branch length (t, in units of changes per codon).

**Estimating patterns of polymorphism in *B. mandarina*.**

We obtained short read sequence data for *B. mandarina* (Xia et al., 2009) from the NCBI short read trace archive (SRP001012). We aligned all data to the reference genome described above. Alignments were performed using BWA (Li and Durbin, 2009) using default parameters. We called genotypes using the GATK (DePristo et al., 2011). We considered only those sites with a minimum of Q30 phred scaled probability of being correctly categorized as either identical to the reference sequence or segregating a non-reference allele. Note that this quantity is computed across the entire sample and individual genotypes may still be relatively low quality, or altogether absent, as the sequencing depth of approximately 3-fold coverage per individual was quite low (Xia et al., 2009). Because the statistics we calculate are concerned only with the number of segregating sites and fixed sites, and not the frequencies of polymorphic variants, this quantity is appropriate for the population genetic analyses we performed. Of the 11 *B. mandarina* individuals with sequence available, 5 are male and 6 are female (estimated from relative Z:A coverage).

Given the inclusion of some female individuals, we observe fewer segregating sites per bp on the Z chromosome than the autosomes (0.0118 vs 0.0229, Mann-Whitney U $P < 2.2 \times 10^{-16}$), which is expected given the expected lower average coverage on the Z and the uniform phred score cutoff we use to call polymorphisms. However, using only males would drastically reduce coverage for polymorphism data, and even if differential coverage has the potential to introduce some bias, there is no reason to expect that bias to differentially affect male-biased and female-biased genes. Thus, we include both males and females in our polymorphism analysis, with the caveat that that overall faster-Z effect in DoS could potentially be influenced by differential coverage (but we do not believe the differential effect of male- and female-biased genes will be impacted).

For those instances in which two or more substitutions were observed within a single codon, we computed the number of nonsynonymous and synonymous changes that are necessary for each possible path and conservatively selected the path that requires the fewest nonsynoymous substitutions, using a custom perl script. Fixed differences were identified as those mutations that are fixed between the *B. mandarina* sample and *B. huttoni*. The reference *B. mori* genome was not used beyond its purpose as an alignment tool.

From the polymorphism tables generated by this procedure, we estimated the Direction of Selection (DoS) for each gene, which is defined as the difference in the proportion of fixed differences that are nonsynonymous compared to the proportion of polymorphisms that are nonsynonymous, and is positive for cases with an excess of fixed replacements, and negative for cases with an excess of polymorphic replacements (Stoletzki and Eyre-Walker, 2010). We use DoS, as opposed to alternative approaches such as the Neutrality Index (Rand and Kann, 1996) or estimating the proportion of fixed amino-acid mutations that have been driven by positive selection (e.g., Welch, 2006), as DoS is much less sensitive to low cell counts than other methods (Stoletzki and Eyre-Walker, 2010).

In addition to DoS, we estimated the ratio of $N_eZ$ to $N_eA$ in *B. mandarina* based on the ratio of mean nucleotide diversities of Z and autosomal chromosomes in the sample. Specifically, we again selected those sites with a minimum sample quality of Q30. We then used only four-fold degenerate synonymous sites to compute mean nucleotide diversity, $\pi$ (Tajima, 1983) on the Z and autosomes. The ratio of these quantities is expected to be the same as the ratio of effective population sizes assuming equal mutation rates of males and females. Changing the minimum quality threshold to Q20 did not significantly affect our estimate. We estimated 95% confidence intervals by bootstrapping.

**Estimating codon bias in *B. mori***

We estimated codon usage bias for *B. mori* sequences for each gene with at least a partial alignment to *B. huttoni*. We used ENCprime to estimate Ncp, the effective number of codons corrected for background sequence composition (Novembre, 2002; Wright, 1990), for each gene. Ncp is a useful measure of codon usage bias, as it does not depend on a defined set of preferred codons, but rather reflects how much codon usage in a gene departs from proportional representation of all synonymous codons under the predictions based on background (non-coding) sequence composition.

**Defining sex-biased genes**

To define sex-biased genes in silkmoths, we relied on published microarray data in *B. mori*, which looked at expression in 9 tissues in both males and females (Walters and Hardcastle, 2011; Zha et al., 2009). Based on the published PTL normalization and model design matrices (Walters and Hardcastle, 2011), we estimated the male/female expression ratio separately in each tissue with the Bioconductor package limma (Smyth, 2005). We define sex-biased genes separately for each tissue as genes expressed significantly differently between males and females at an FDR of 10% and with a fold-change of expression at least 1.5x. To combine data across all tissues, we consider as male-biased any gene that is male-biased in at least one tissue but not female-biased in any tissue, and vice versa for female-biased genes. Genes which are female-biased in some tissues and male-biased in others are considered unbiased, along with genes that are not sex-biased in any tissue. Overall, there are 1,414 female-biased genes, 4664 unbiased genes, and 2082 male-biased genes. Of these, 59, 190, and 138 are on the Z chromosome, respectively, with the remainder on autosomes.

**Statistical analysis**

After estimating evolutionary parameters for each gene, we performed most statistical analysis in R version 3.0.0. In general, we are interested in comparing medians of distributions between autosomal and Z-linked genes, which we do using approximate Wilcox-Mann-Whitney tests that use 100,000 Monte Carlo resamples to calculate P-values, as implemented in the function wilcox_test from the R package coin. In order to estimate ratios of medians and confidence intervals, we use a weighted bootstrap (ordinary importance resample), implemented in the R package boot and using 10,000 bootstrap replicates. We calculate the median ratio as the mean of the bootstrap resamples, and the 95% confidence interval using the "percentile" method in the R function boot.ci, unless otherwise indicated. Prior to analysis we scaled DoS to be strictly positive by adding 1 to each value, in order to make the median ratio interpretable. To test differences in the median ratios between male-biased and female-biased genes, we used a permutation test in which the chromosome and sex-bias assignments for each gene were randomly permuted 10,000 times; for each permutation we calculate the difference in Z/A median ratios between male-biased and female-biased genes to generate a null distribution on this statistic.

**Acknowledgments**
We would like to thank Nadia Singh, James Walters, and Rich Meisel for helpful discussion and comments on the manuscript. This work was supported by NIH grants GM084236 and GM065169 to DLH, and by the Department of Biotechnology, Government of India through Task force grant to KPA and JN.

**Author contributions**
Designed the experiment: TBS JN DLH. Collected data: TBS RBCD RLV KPA JN. Analyzed data: TBS RBCD. Wrote the paper: TBS RBCD DLH.

**Figure legends**

**Figure 1. A)** Distribution of contig lengths in the final assembly. The dashed line indicates the N50 value. **B)** Aligned coverage of *B. mori* genes based on unique promer mappings.

**Figure 2.** Boxplot of ω, dN, and dS (left to right) for autosomal and Z chromosome genes in *B. mori* / *B. huttoni* alignments. Median ω and median dN are significantly different between chromosome classes (P = 3 x $10^{-5}$ and P = 7.3 x $10^{-4}$, Wilcox-Mann-Whitney test), but dS is not (P = 0.1545).

**Figure 3.** Faster-Z effect in male-biased, unbiased, and female-biased genes. (left) The faster-Z effect is Z:A ratio of median ω, on a log2 scale, corrected for differences in alignment coverage using a weighted bootstrap. Error bars represent 95% confidence intervals from the weighted bootstrap. The value for female-biased genes is significantly greater than the value for male-biased genes based on a permutation test (P = 0.0124). (right) The faster-Z effect is Z:A ratio of median scaled DoS (transformed by adding 1 so that all values are positive and to improve stability of bootstrap estimates), on a log2 scale, weighted by the DoS.weight parameter (Stoletzki and Eyre-Walker, 2010) using a weighted bootstrap. Error bars represent 95% confidence intervals from the weighted bootstrap. The value for female-biased genes is significantly greater than the value for male-biased genes based on a permutation test (P = 0.005).

**Appendix**

In order to calculate the expected impact of background selection on $N_eZ:N_eA$ at neutral sites in *B. mori*, we start from the results derived by Charlesworth (2012b) for the overall effect of background selection on levels of variability (equation 5b in the referenced paper), which states that:

$B \approx \exp(-U/M)$

where B is the effect of background selection, U is the deleterious mutation rate per chromosome, and M is the population effective map length.

The population effective map lengths are easily calculated from the published linkage map (Yamamoto et al., 2008), which implies an average male autosomal map length of 0.50 M and a Z map length (in males) of 0.45. To convert these to population effective map lengths, the autosomal length is multiplied by 1/2 and the Z length by 2/3, giving 0.25 M and 0.3 M, respectively. Based on the *B. mori* genome, we can calculate the fraction of the genome associated with the Z and with the average autosomal arm as 0.045 and 0.0354, respectively.

There is no estimate of U for silkmoths, so for simplicity we assume a value of 1, which is often used for *D. melanogaster* (e.g., Charlesworth, 2012b), but assuming a range of U values produces identical results. The predicted B values for autosomes and the Z chromosome are respectively 0.861 and 0.868, with a Z:A ratio for B of 1.01, giving an expected Z:A ratio of 1.01 x (3/4), or 0.756. Clearly, differential effects of background selection have little effect on the expected neutral diversity ratio for the Z and the autosome in this species.

**Figure 1A:**

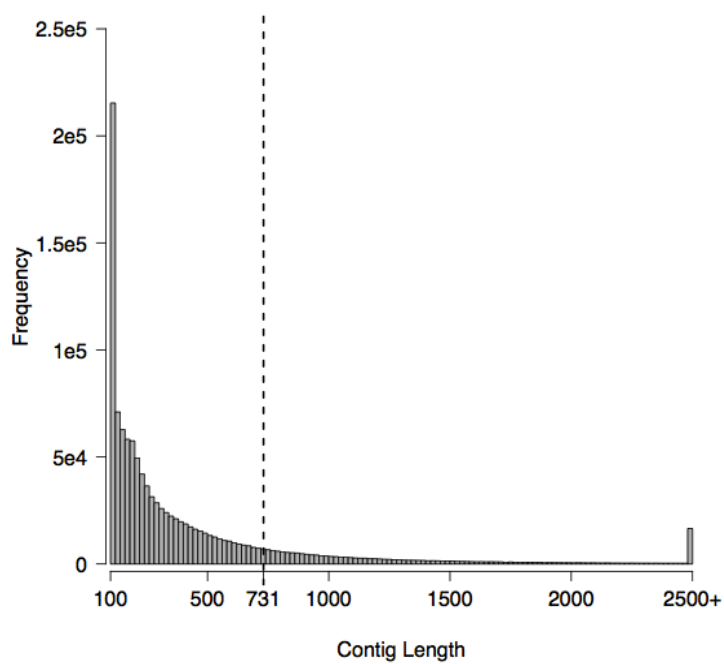

**Figure 1B:**

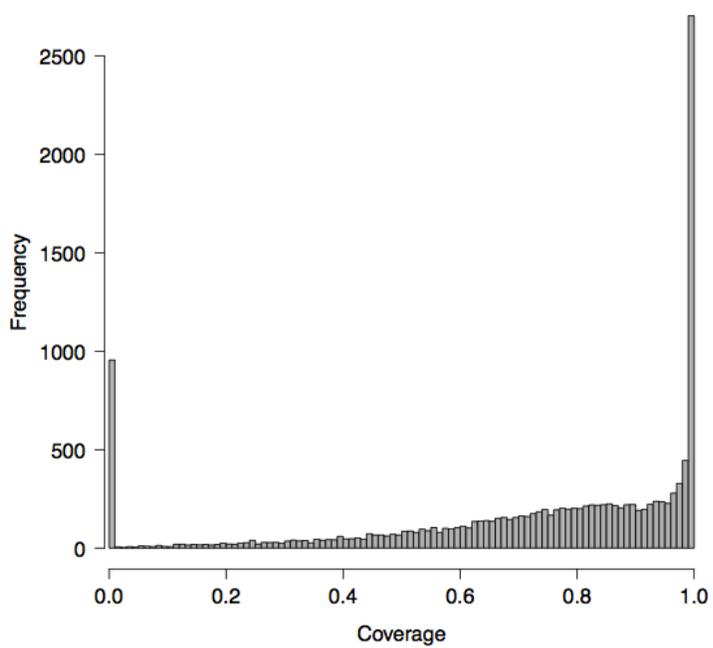

**Figure 2:**

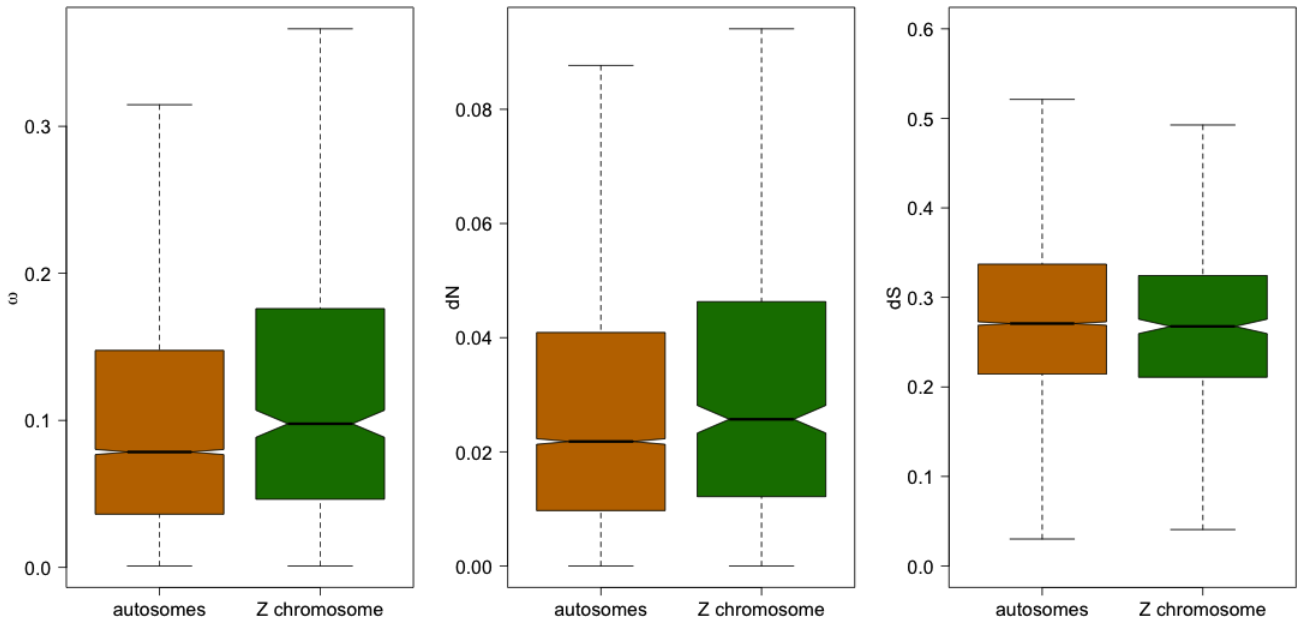

**Figure 3:**

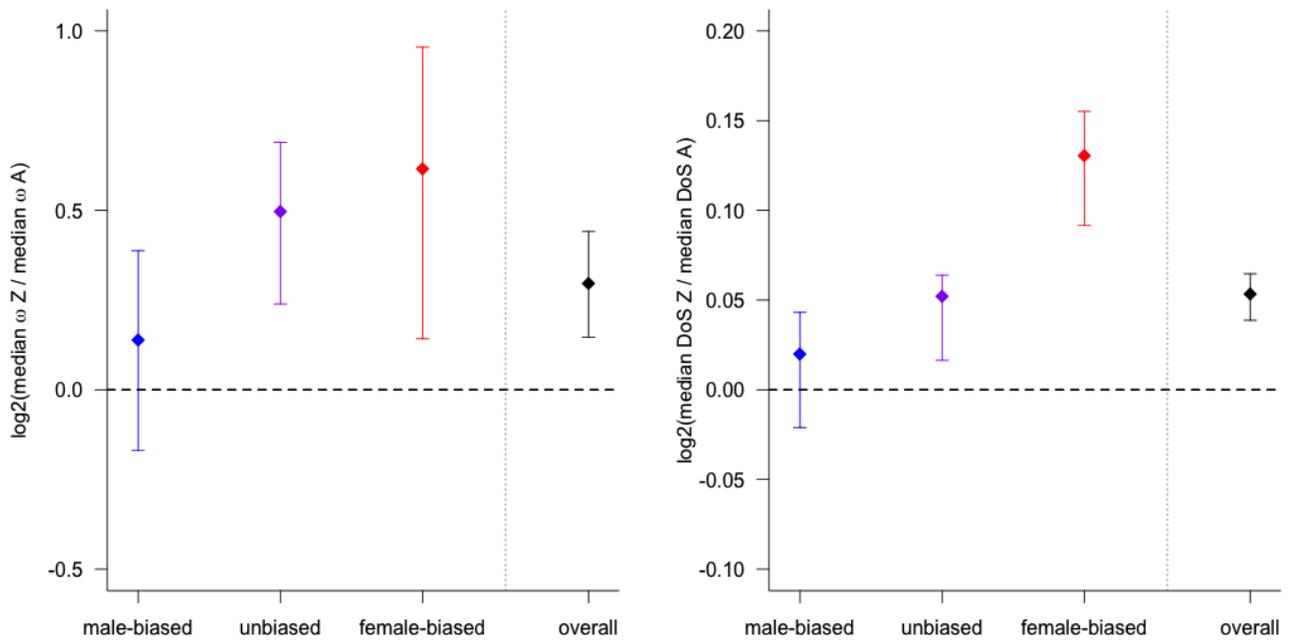